# MODELLING DIFFUSE SUBCELLULAR PROTEIN STRUCTURES AS DYNAMIC SOCIAL NETWORKS

by

ANDREW MACLEAN DURDEN

(Under the Direction of Shannon Quinn)

ABSTRACT

Fluorescence microscopy has led to impressive quantitative models and new insights gained from richer sets of biomedical imagery. However, there is a dearth of rigorous and established bioimaging strategies for modeling spatiotemporal behavior of diffuse, subcellular components such as mitochondria or actin. In many cases, these structures are assessed by hand or with other semi-quantitative measures. We propose to build descriptive and dynamic models of diffuse subcellular morphologies, using the mitochondrial protein patterns of cervical epithelial (HeLa) cells. We develop a parametric representation of the patterns as a mixture of probability masses. This mixture is iteratively perturbed over time to fit the evolving spatiotemporal behavior of the subcellular structures. We convert the resulting trajectory into a series of graph Laplacians to formally define a dynamic network. Finally, we demonstrate how graph theoretic analyses of the trajectories yield biologically-meaningful quantifications of the structures.

INDEX WORDS:    Biomedical Imaging, Graph Theory, Social Networks, Gaussian Mixture Modeling, Mitochondria, Subcellular Proteins

MODELLING DIFFUSE SUBCELLULAR PROTEIN STRUCTURES AS DYNAMIC

SOCIAL NETWORKS

by

ANDREW MACLEAN DURDEN

A Thesis Submitted to the Graduate Faculty of The University of Georgia in Partial

Fulfillment of the Requirements for the Degree

MASTER OF SCIENCE

ATHENS, GEORGIA

2019

MODELLING DIFFUSE SUBCELLULAR PROTEIN STRUCTURES AS DYNAMIC

SOCIAL NETWORKS

by

ANDREW MACLEAN DURDEN

| | |
|---|---|
| Major Professor: | Shannon Quinn |
| Committee: | Tianming Liu |
| | Frederick D. Quinn |

Electronic Version Approved:

Suzanne Barbour
Dean of the Graduate School
The University of Georgia
May 2019

# DEDICATION

I would like to dedicate this work to Caroline Woodring. It would not have been completed without your support and encouragement.

ACKNOWLEDGEMENTS

I would like to thank Mojtaba Fazli and Shannon Quinn for their help, work, and guidance on this project.

I would also like to thank Allyson Loy, Barbara Reaves, Abigail Courtney, Chakra Chennubhotla, Frederick Quinn, Brittany Dorsey, and Chinasa Okolo. Their work with the Ornet project has made my research possible.

This project was supported in part by a grant from the National Science Foundation (#1458766).

TABLE OF CONTENTS

LIST OF FIGURES

CHAPTER 1

INTRODUCTION

Significant portions of the material in this thesis has been published in the proceedings of the Scipy Conference 2018. Listed below is the publication citation and the chapters of this thesis which reference it.

*Dynamic Social Network Modeling of Diffuse Subcellular Morphologies.* Andrew Durden, Allyson T Loy, Barbara Reaves, Mojtaba Fazli, Abigail Courtney, Frederick D Quinn, S Chakra Chennubhotla, and Shannon P Quinn. Published in the proceeding of the Scipy Conference 2018. Doi 10.25080/Majora-4af1f417-000

- CHAPTER 1 INTRODUCTION – Background, Data
- CHAPTER 2 SEGMENTATION – Our Segmentation Pipeline
- CHAPTER 3 SOCIAL NETWORK MODEL – Our Network Model
- CHAPTER 4 MODEL ANALYSIS AND CONCLUSION

Background

Optical imaging has been advancing steadily for many years; with it, our ability to visually monitor biological phenomena has increased in scale, specificity, and complexity (Eliceiri et al. 2012). As a result, biologists grew increasingly interested in analyzing images to convert microscopy images into quantitative data (Eliceiri et al. 2012). In the early 2000s there was a heavy effort in modeling the localization of subcellular proteins. Yet, at this point the primary means of analyzing and annotating images of was manual visual examination (Glory and Murphy 2007). Many large sets of manually annotated microscopy videos and images were created including a set of 40,000 movies of early embryogenesis in *Caenorhabditis elegans* (Ljosa & Carpenter 2009). As automated microscopes proliferated in use, image volume became overwhelming for manual annotation and analysis. In turn scientists began to employ automated analyses for speed, conducting experiments on a much grander scale than previously. While reduction of human labor was a common goal of automated image analysis, new models were able to yield objective and quantitative measurements of relevant biological phenomena, sometimes even distinguishing differences manual annotation failed to notice (Ljosa & Carpenter 2009).

Early efforts in quantitative analysis of microscopy images focused on segmenting objects like nuclei and cells. This was done with intensity thresholding, splitting clusters through the watershed algorithm, and creating simple measurements such as counts, size, shape, and localization (Ljosa & Carpenter 2009). These techniques began democratizing automated image analysis and image processing became more commonplace. Many generalist image analysis tools became more readily available in the following years, such as BioImageXD (Kankaanpää P, et al. 2012), Icy (de Chaumont et al. 2012), and Fiji (Schindelin et al. 2012), among others.

In the last decade, deep learning has introduced new paradigms for conducting image analysis. Convolutional neural networks (CNNs), a specific instantiation of deep learning well suited for image analysis, gained a large amount of momentum as new, efficient training algorithms were developed. In the years following AlexNet's (Krizhevsky et al. 2012) victory in the ImageNet challenge in 2012, CNNs became ubiquitous in the field of computer vision. The medical community has, gradually, begun to apply deep models to studies in their field (Litjens et al. 2017). One of the most recent advancements in the field of deep medical image annotation is the U-net (Ronneberger et al. 2015) which is still widely considered the state of the art for medical image segmentation.

However, much of this advancement has been in the context of a narrow and well-behaved subset of biological morphologies. The democratization efforts in Ljosa & Carpenter 2009 discusses "objects" like cells and nuclei, presenting techniques to measure the size and shape of these contiguous structures. Even more recent applications of deep models like Xie et al. 2018 focus on counting "solid" cell bodies. These "solid" objects are relatively simple to track; however, these models fail when applied to biological structures with diffuse patterns such as those created by mitochondria or actin tubules.

## Significance and Motivation

Developing a quantitative model for the spatiotemporal behavior of subcellular organelles' responses to external stimuli will advance our understanding of the stimuli's effects on the internal state of the cell. These advances have implications on the production of therapies targeted at these specific stimuli. There has been recent work in measuring covariance in organellar response to external stimuli with the goal of better understanding cellular dynamics (Valm et al. 2017). While this work featured a heavy amount of manual intervention, other recent work has resulted in an open source pipeline for measuring spatial covariance in biological data, specifically in gene expression (Svennson et al. 2017). These two illustrate the purpose of our work: to develop an open pipeline for the generation of quantitative spatiotemporal models of diffuse subcellular structures.

It is known that mitochondria also have a strong relationship with cell morphology. A recent study sought to better model mitochondria and characterize the relationship between cell morphology and mitochondrial distribution. In this study, a gaussian mixture model was fit to a three-dimensional image of mitochondria (Ruan et al. 2018). This study independently replicates our approach to modeling mitochondria; however, our process aims to describe the dynamic temporal behavior as well as the spatial distribution of the mitochondrial protein structure.

Unlike many of these previous studies, our goal is not only to model the spatial covariance of the subcellular structures, but to understand how these structures respond spatially to stimuli over time. With these goals in mind, we leverage graph theory and social network analogues to model and analyze our subcellular structures. A recent study in brain imaging made use of graph theory to create a quantitative measure of activity in functional MRI images for unsupervised learning (Drysdale et al. 2016). A graph-based approach has been applied recently to deep learning on point clouds. Neighborhood graphs applied to the point clouds provide a quantification of the structure described by many points suspended in 3-dimensional space (Wang et al. 2018). Using social networks to model certain phenomena is useful to capture both the overall morphology of the structure being modeled as well as the relationships between various parts of the structure. We propose to adapt of these graph theoretic constructs to model diffuse subcellular proteins, capturing both the spatial distribution of the proteins as well as the spatial relationships across the structure. We then incorporate the temporal evolution of the network, allowing us to interpret the changing network properties in terms of biologically relevant phenomena. Studies have shown methods of empirically assessing global effects of local phenomena in networks over time (Kossinets & Watts 2006). The evolution of our fitted network models serve as a model of the behavior of the mitochondria as the cell responds to various stimuli.

For the purposes of this study, we focused on modeling the subcellular patterns of mitochondria in cervical epithelial (HeLa) cells. Mitochondria are dynamic organelles, which undergo continual rounds of fission and fusion. These fission and fusion events are important for maintaining proper function and overall mitochondrial health (Zhao et al. 2013, Wai et al. 2016). Mitochondrial fission allows for the turnover of damaged organelles and the protection of healthy ones. Additionally, mitochondrial fusion leads to the mixing of internal contents, such as mitochondrial DNA, which is important for responding to environmental needs (Zhao et al. 2013, Knott et al. 2009). The dynamics between fission and fusion create a spectrum of mitochondrial morphologies. Imbalances between fission and fusion events generate morphological phenotypes associated with mitochondrial dysfunction (Zhao et al. 2013). An excess of fission or dearth of fusion events results in fragmented mitochondria; in this phenotype, the mitochondrial phenotype appears fractured, and individual mitochondria exist in small spheres. Conversely, an overabundance of fusion or a lack of fission events generate hyperfused mitochondria; in this phenotype, the mitochondrial structure appears overconnected, composed of long interconnected tubules (Cassidy-Stone et al. 2008). Recently, several bacterial species have been shown to cause mitochondrial perturbations during infection (Stavru et al. 2011, Fine-Coulson et al. 2015). Such unique morphologies should be detectable at a quantitative level using social network modeling.

Through social network modeling, we hope to build a more rapid and efficient method for identifying changes in size, shape, and spatial distribution of mitochondria as well as other diffuse organelles. Figure1 shows our pipeline, which begins by segmenting cells with fluorescent markers on the mitochondria. Once the cells are segmented, we use a Gaussian Mixture Model (GMM) to parameterize the spatial distribution of the mitochondrial protein patterns at evenly-spaced time intervals and allow the GMM parameters to update smoothly from the previous time point to the next. Finally, we demonstrate how the learned parameters of the GMM can be used to construct social networks for representing the mitochondria.

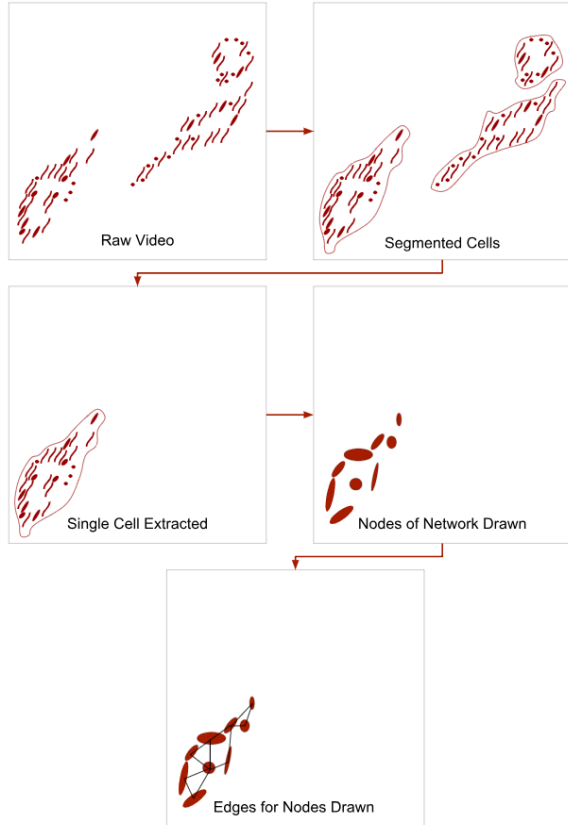

Figure 1. Abstract Representation of Proposed Pipeline: We begin in the top left with the raw image data when then follow the arrows in the figure. First, we segment cells. Second, we extract each cell. Third, we apply a mixture model to generate nodes. Finally, we add edges to the network to complete the network structure.

Dataset

We have constructed a library of live confocal imaging videos that encompass a spectrum of mitochondrial morphologies in HeLa cells, from fragmented to hyperfused. To visualize the mitochondria, HeLa cells were stably transfected with DsRed2-Mito-7 (DsRed2-HeLa), which fluorescently labels mitochondria with red emission spectra (a gift from Michael Davidson, Addgene plasmid #55838). All of the videos were taken using a Nikon A1R Confocal. Cells were kept in an imaging chamber that maintained 37 degrees C and 5% $CO_2$ for the duration of imaging. The resonant scanning head was used to capture an image every ten seconds for the length of the video. The resulting time series videos have more than 20,000 frames per video. Each frame is of dimensions of 512x512 pixels (Figure 2).

Wild type mitochondrial morphology was captured by imaging DsRed2-HeLa cells in typical growth medium (DMEM plus 10 % fetal bovine serum) (Figure 2, center). To generate the fragmented phenotype, cells were exposed to the pore-forming toxin listeriolysin O (LLO) at a final concentration of 6 nM (Figure 2, left). Mitochondrial hyperfusion was induced through the addition of mitochondria division inhibitor-1 (mdivi-1) at a final concentration of 50μM (Figure 2, right). These subsets with different known qualitative phenotypes serve as bases upon which to condition our quantitative analyses.

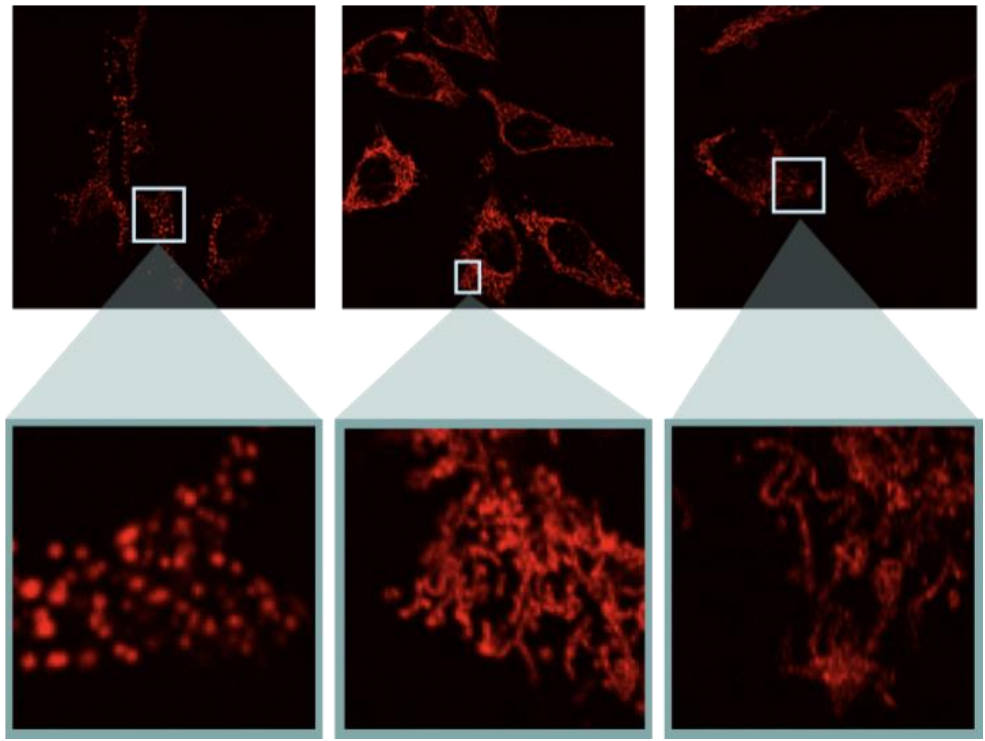

Figure 2. Mitochondrial Morphologies: (Left) LLO induced mitochondrial fragmentation. (Center) Wild type HeLa mitochondrial morphology. (Right) Mdivi-1 induced mitochondrial hyperfusion.

CHAPTER 2

SEGMENTATION

In order to avoid systemic bias in our downstream analysis as a result of different videos containing a varied and unbounded number of cells, we chose to study each cell individually. This required segmenting each cell and studying its spatiotemporal dynamics in isolation from the others. While segmentation of cells from fluorescence or histology images is a common operation, segmenting diffuse protein patterns—such as mitochondria—is much more challenging.

### Thresholding

One of the most well-established techniques for separating objects in an image from their background are thresholding methods. Thresholding is a simple yet powerful tool, as the image background is often uniformly dark while the object of interest fluoresces brightly (Ljosa & Carpenter 2009). There are numerous techniques for thresholding images which have all been consistently applied to solid body imaging. Thresholding methods are varied in approach and are commonly split into two categories: locally adaptive thresholds and global thresholds. In locally adaptive thresholds each pixel has its own threshold calculated based on its neighboring pixels. This is done by determining the threshold of a pixel at ($i,j$) as some function T($i,j$). An early approach to the T($i,j$) function was to compute the local mean and standard deviation of pixel intensities in a neighborhood of some size $b \times b$ and adapt a threshold value based in these metrics (Sezgin and Sankur 2004). This base approach has worked well in binarizing text documents in poorly illuminated images (Sauvola & Pietikäinen 2000). Another classical adaptive thresholding method fits a surface to the grey-levels in a single channel image. This approach works by modeling the background of the image as a surface which mimics any change in background intensity over the spatial region and uses this surface as the threshold to binarize the image with separate thresholds at each pixel (Sezgin and Sankur 2004). A different version of this method uses a variational approach to reduce the number of parameters to one, while preserving the effectiveness of the surface-fitting approach (Chan et al. 1998). Locally adaptive thresholds are most effective in images in which lighting or intensity vary across the spatial dimensions of the data. Consequentially, due to the consistent background in our data set we focused on global thresholds.

Global thresholding techniques use a single threshold value for the entire image. This value can be manually set for a well-behaved dataset or computed as a function of the image data. There are several methods by which the global threshold can be determined. One approach to global thresholding focuses on the shape of a smoothed histogram of greyscale values. Features describing the shape of a grey-level histogram are used to determine a global threshold. Other processes use the convex hull to find concavities in the smoothed histogram. These concavities are considered possible thresholds; variations on the convex hull approach use different methods to further determine the best of the possible thresholds (Sezgin and Sankur 2004). Another common approach is to find the peaks and valleys in the grey-level histogram. To search for the peaks and valleys the smoothed histogram is differentiated, and zero crossings are found in the result. A triple of zero crossings give an initial, maximum, and final zero crossing in which an optimal threshold lies. Various approaches are used to find the optimal peak-valley triplet in which the threshold is between the initial and final zero crossing (Sezgin and Sankur 2004). A similar set of algorithms cluster the grey level histogram into two separate groupings representing the object or the background. The most commonly used form of this approach is the Otsu thresholding method (Otsu 1979). Otsu's method uses an exhaustive search through the grey values to determine a threshold with the minimal intra-class variance. By minimizing intra-class variance, the method in turn maximizes the scatter between the object and background clusters (Otsu 1979). A similar approach is ISODATA clustering. This iterative approach was first discussed in a study determining a picture segmentation threshold. In the study, the proposed method gives a threshold which separates the image into two groups of pixels with a value halfway between the means of the two groups (Ridler & Calvard 1978). A later study linked this approach the known ISODATA clustering algorithm applied in a single dimensional feature space: grayscale pixels. Because of the lack of background intensity variation in our data and the effectiveness of these last two methods we generally used Otsu and ISODATA clustering in our segmentation work.

Common post-processing steps when using threshold based segmentation include dilation to grow the region of interest past the threshold boundaries, and a closing technique or convex hull to create a solid mask. We followed this approach, using ISODATA thresholding on a frame with many cells, applying dilations and closing to expand the area and remove small objects, and computing a convex hull to create masks. Our initial results were inaccurate, occasionally merging nearby cells, but more often splitting cells into separate masks (Fig 3). Because of this result when using traditional threshold-based segmentation on the diffuse and nearly overlapping objects, we "primed" our segmentation procedure with hand-drawn masks at time 0. This allows us to leverage the fact that, given the small interval (10s) between frames of a video, overall movement between a given pair of subsequent frames would be small, permitting the use of deformable contours.

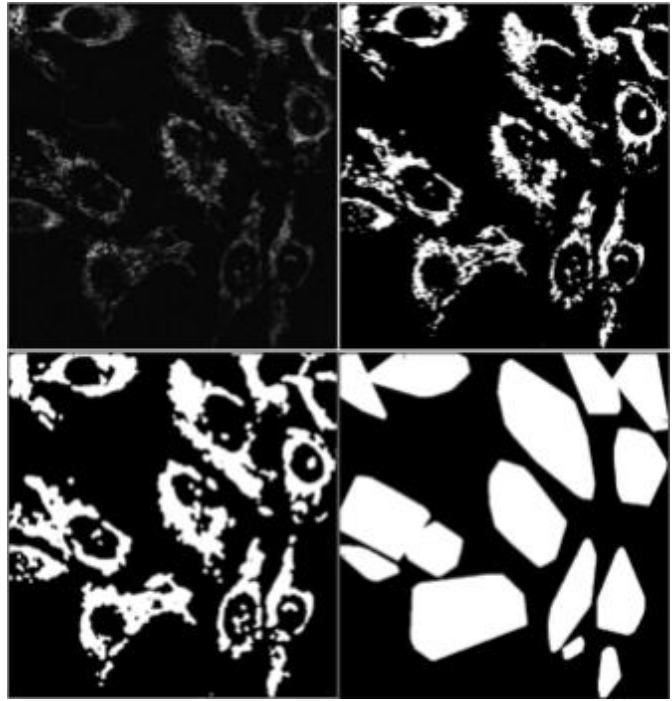

Figure 3. Thresholding Results: (Top Left): raw frame from video containing large number of cells. (Top Right): result of ISODATA threshold. (Bottom Left): after applying dilation post-processing. (Bottom Right): final masks after convex hull. Notice the separation of pixels from the same cells in the bottom right of the image as well as cell merging in the top right of the image.

<u>Deformable Contours</u>

We used the ITK-SNAP software (Yushkevich, et al. 2006) to manually draw segmentation boundaries for each cell in the first frame of each video, generating a VTK file with the segmentation maps (Fig. 7, top left). To deform the masks, we initially looked to using a snake active contour model (Kass, et al. 1988) to deform our hand drawn contours over time. The snake uses an energy-minimizing spline to accurately localize edges by conforming to a gradient around an object and make iterative adjustments until it reaches an energy equilibrium point. Snakes are able to adjust to small movements between frames for object tracking. Due to the low motion nature of our data we hoped that the snakes would both reduce overlaps in our masking and provide a method for tracking the shift in mitochondrial protein patterns between each frame. We applied the snakes active contour model to automatically evolve our initial hand-drawn masks over the full duration of each video. The results of this model were overly conservative, often finding edges between the protein's main structure and smaller 'rogue' clusters in a single protein pattern (Fig 4). Another issue we encountered was processing time. The snakes model's convergence on an equilibrium was computationally intractable and scaled poorly to larger data.

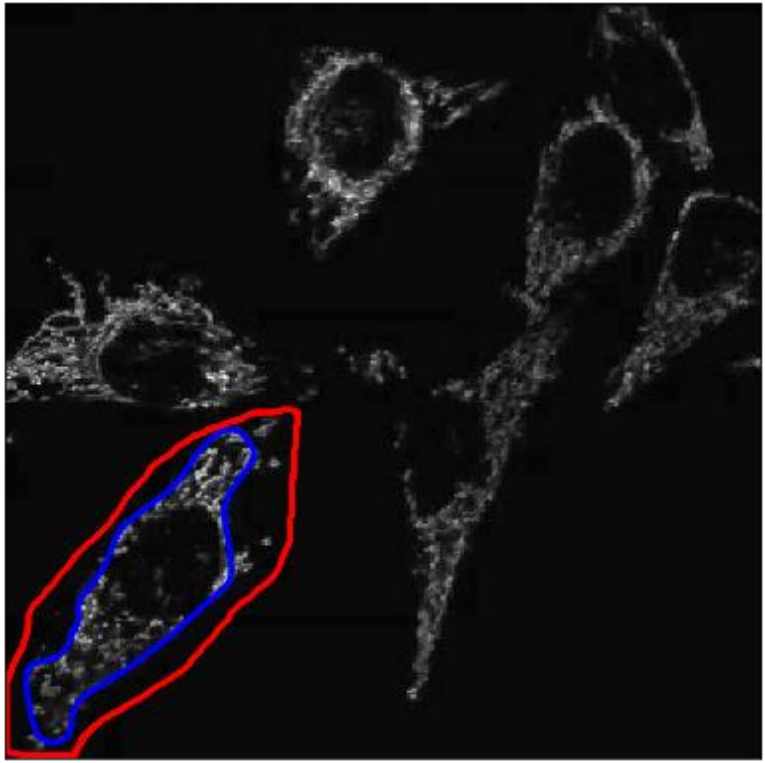

Figure 4. Snakes Contour: hand drawn mask boundary shown in red and the converged snake active contour shown in blue. Notice the 'rogue' mitochondrial clusters accounted for in the red bounds which are lost in the blue bounds.

We developed a custom process which merged the deformable contour with a classical thresholding approach using Otsu's method. Like snakes, our custom segmentation process used our hand drawn VTK masks as "seeds" and updated the maps at each frame of the video using our own deformation process of "in mask" thresholding, dilation, and overlap detection processes over the entire video (Fig 5). This process resulted in a set of masks for each frame, one for each cell in the frame, which could be used to extract individual cells over the course of the video (Fig. 7).

```
Algorithm 1 Segmentation Algorithm
Input: A video $V \in R^{h \times w \times f}$; A set of initial binary mask images $M \in R^{h \times w \times c}$;
c, The number of distinct cells
Output: A set of generated contour masks for each frame $ContourList \in R^{h \times w \times c \times f}$
 1: function CREATECONTOURS(V, M, c)
 2:     $ContourList \leftarrow \emptyset$                    ▷ initialize output with empty set
 3:     for i = 0 to f do                    ▷ for each frame in video generate contours
 4:         $F \leftarrow V_i$
 5:         $Contours \leftarrow $ FITTIGHTCONTOURS(F, M, c)
 6:         $Contours \leftarrow $ DILATIONWITHOVERLAPDETECTION(Contours)
 7:         $M \leftarrow Contours$           ▷ Use current contours as seed for next frame
 8:         ContourList.APPEND(Contours)
 9:     return ContourList                                      ▷ return all masks
10:
11: function FITTIGHTCONTOURS(F, M, c)
12:     $Contours \leftarrow \emptyset$                       ▷ initialize output with empty set
13:     for i = 1 to c do                                         ▷ For each cell
14:         $Contours_i \leftarrow F$                           ▷ Copy raw frame data
15:         $Contours_i[M_i \neq 1] \leftarrow 0$         ▷ remove raw data outside of seed
16:         $Contours_i \leftarrow threshold(Contours_i)$  ▷ Convert to binary mask
17:         $Contours_i \leftarrow close(Contours_i)$     ▷ remove interior holes
18:     return Contours
19:
20: function DILATIONWITHOVERLAPDETECTION(Contours)
21:     $Contours_i \leftarrow dilate(Contours_i)$   ▷ dilate to get mass outside seed
22:     for i, j = 0 to length(Contours) do               ▷ Do pairwise comparison
23:         $inter \leftarrow Contours_i \cap Contours_j$
24:         if $inter \neq \emptyset$ then                     ▷ If cells overlap, cut overlapping area
25:             $Contours_i \leftarrow Contours_i$ xor $inter$
26:             $Contours_j \leftarrow Contours_j$ xor $inter$
27:     return Contours
```

Figure 5. Segmentation Algorithm Pseudocode

<u>Our Segmentation Pipeline</u>

Our algorithm first fit tight masks by applying a threshold to the area inside of the seed masks (or the masks generated from frame t-1) as a tight deformation (Fig 5 line 11). While this process was very effective at following the cells over the video, occasionally our contours would lose small areas of mitochondrial mass which were sufficiently far away from the more contiguous structure, similar to the loss found in the snakes approach and the losses found in the application of a classic thresholding based segmentation. To compensate, we added a process of iterative dilation to prevent loss and give a more generous contour (Fig 5 line 20). This allowed our version of the deformable contour to be less focused on fitting visible lines tightly and more focused on clustering groups of non-contiguous but nearby proteins together into their containing cell.

With these adjustments, we ran into a rare problem of cell contact or overlap. We altered the iterative dilation to perform more iterations with smaller dilation kernels, repeatedly checking for overlap with any other map on each iteration (Fig 5 line 24). In the case of an overlap, which would only be a few pixels with the small dilation kernel, we used a simple XOR to remove the few overlapping pixels while still allowing the mask to expand in areas unclaimed by other cells (Fig 6). With this case being rare, we found the process mostly followed any visible boundary of the adjacent cell.

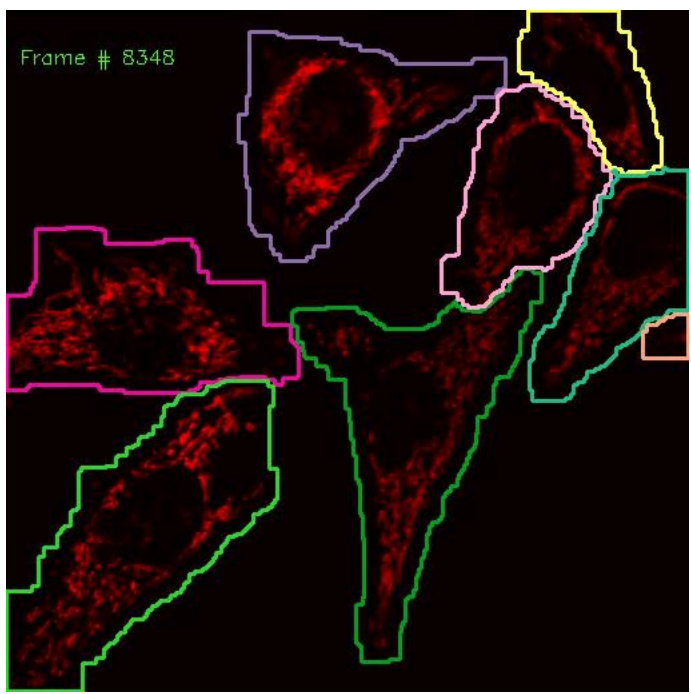

Figure 6. Segmentation Results: The middle frame of a crowded video during segmentation. Notice the generous boundaries along uncontested edges of a cell (along the bottom regions of the dark green contour) and tight borders where cells become close (the rightmost edge of the pink contour).

Once the pairwise overlap checking was implemented, we faced runtime issues similar to the snakes approach. In videos containing >10 cells (Fig 6), a pairwise intersection on every iterative dilation became computationally expensive. To improve our runtime, we leveraged the low motility of the cell bodies in our dataset. As our cells were nearly immobile, two cells were unlikely to begin to overlap later in the video. As protein structures fragmented, they were likely to cease overlapping with previous neighbor cells. Accordingly, we created a lookup list in the first twenty "priming" frames of each pair of cell masks which intersected during dilation. After these priming frames we only compared those pairs in the list. We continued to update the list every 100 frames in case of cell movement, or fragmentation which may have caused two cells to begin to or cease overlap. These adjustments allowed us to quickly segment our entire video dataset with minimal protein loss, allowing us to rapidly extract each individual cell as a distinct video so the network model could be agnostic to the number of cells (Fig 7).

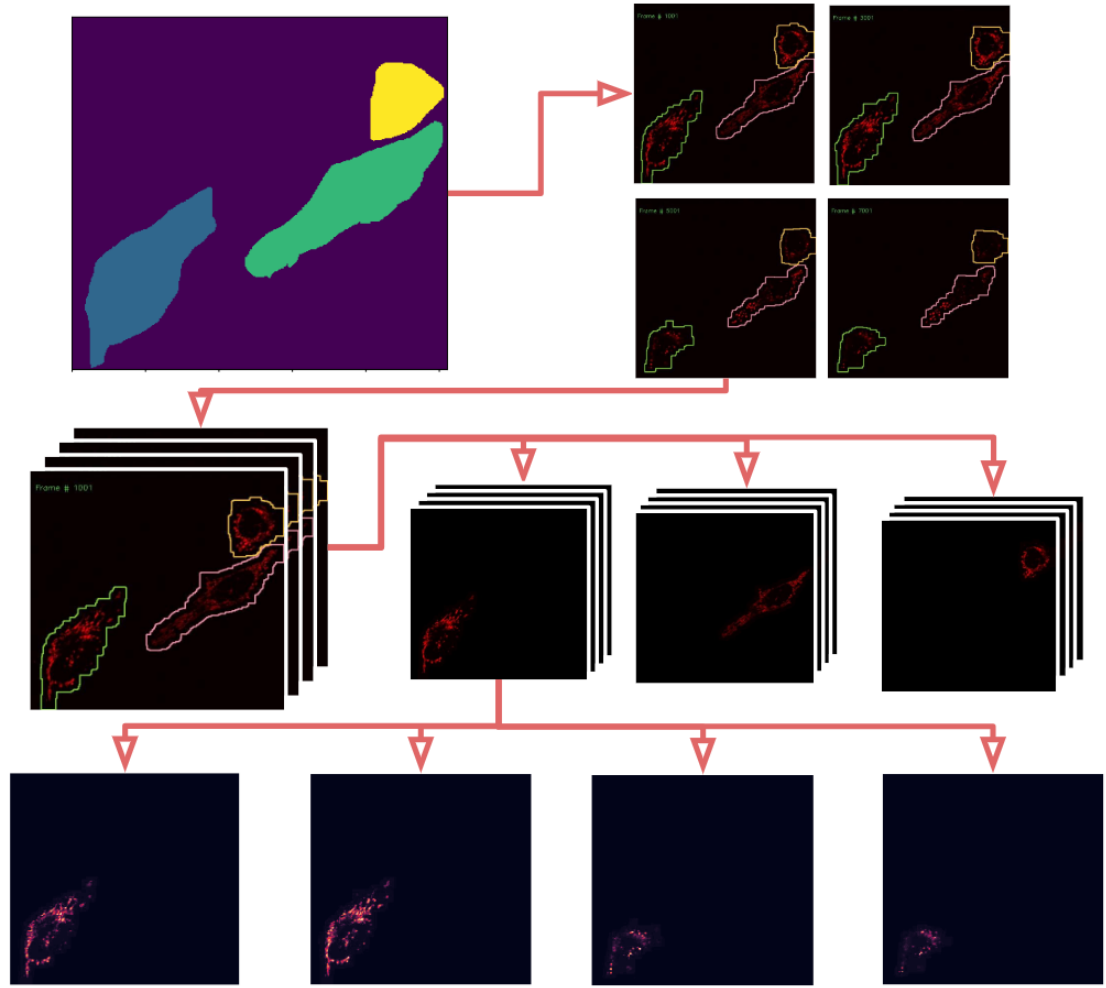

Figure 7. Cell Segmentation Pipeline. (Top Left): Hand drawn masks of the first frame in VTK format were used to "seed" the deformable contours. (Top Right): A series of frames from a single video with autonomously drawn contours. (Middle): Stack of frames from a single video converted to separate videos for each cell. (Bottom): single cell video unraveled as grayscale image for frame by frame network modeling.

CHAPTER 3

SOCIAL NETWORK MODEL

Social Networks

Social networks have been studied for decades, initially as an area of sociological work to study people (Scott 1988). Individuals were linked together through some sort of bond or affinity. Many individuals linking together created a crisscross mesh of connections. Previously, these models were used to study kinship structures, social mobility, academic citations, criminal or deviant activity, and other areas of social interactivity (Scott 1988). Public notions of direct connectedness between people gave rise to the adage that any two individuals have at most 6 degrees of separation, stemming from an early study which showed an average of 5.2 degrees of separation amongst the subjects (Milgram 1967). This gave rise to modern "memes" like the Kevin Bacon number, an actor's degree of separation from Kevin Bacon in a social network model, connecting those who have appeared in the same film (Hopkins 2004). Social network models have also been applied to studies in fields outside of sociology, including ecology, internet topology, as well as many studies in microbiology (Strogatz 2001).

Ecology studies have approached food webs with social network frameworks to study how various species consume each other. In these studies, two species in the network are connected if the one consumes the other. This approach outperformed the previously popular cascade model in estimating the empirical food web properties in several varying habitats (Williams & Martinez 2000).

Social Networks have been to web pages for studying the topology of the internet. One study fit a network to static web pages connecting a page to another if the prior page contains hyperlinks to the latter page. This model was used to design web crawling strategies, study content creation, and to predict the emergence of new topological phenomena on the web graph (Broder et al. 2000).

In microbiology social network models are fit to many biological structures to model their interactions. One study built a social network to model metabolic networks, which modeled substrates as the individuals in the network connected to each other by links representing metabolic reactions. The analysis established new characteristics of the structure of metabolic networks and quantified the metabolic network of organisms in all three domains of life: eukaryote, bacterium, and archaea (Jeong et al. 2000). We hope to adapt this style of model to our dataset to leverage graph theoretical concepts to similarly quantify biological phenomena.

### Graph Theory

Social networks model some entity (people, species, web pages, etc.) and their interactions as vertices an edges in a graph. A graph can be defined as a set of nodes or vertices $V$ and a set of edges $E$. Each edge in $E$ connects a pair of nodes $(u,v)$ such that $u$ is connected to $v$. A graph can be directed (Fig 8 b), where $(u,v) \in E \nRightarrow (v,u) \in E$, or it can be undirected (Fig 8 a), where $(u,v) \in E \Rightarrow (v,u) \in E$. An example of a directed graph is the food web network above, where one species eating another links the former to the latter, however this relationship does not link the latter to the former unless the latter species also eats the former species. An undirected graph similar to the food web would be one where species are linked if they co-inhabit an ecosystem as this is a directionless relationship. Many popular social media networks have asymmetric connections between users. For example, Twitter and Instagram permit users to follow another without being followed back. Even Facebook, which has a symmetric "friend" connection, has asymmetric underlying weights in terms of how friends interact over the network.

Graphs can also have edges that are unweighted or weighted (Fig 8 c). A weighted graph has a matrix $W$ such that $W(u,v)$ is the weight or strength of the connection between nodes $u$ and $v$. A weighted graph in social network modelling allows for more complex relationships between the entities being studied by considering the relative strength of interaction between entities. In an unweighted graph, this weight matrix would simply contain ones representing a connection, or zeros representing the lack of a connection. Graphs can be dense or sparse, ranging from connections between every node (complete or fully-connected, Fig 8 d) or just a few connections throughout the graph as a whole.

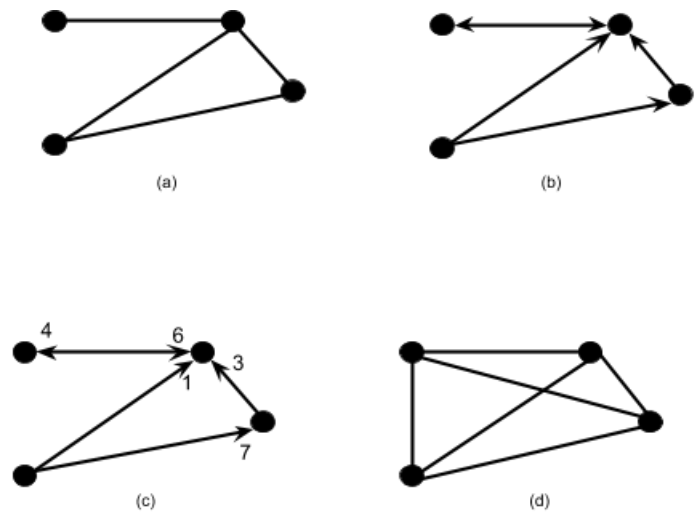

Figure 8. Basic graph types: (a) a partially connected, undirected, and unweighted graph. (b) a partially connected, directed, and unweighted graph. (c) a partially connected, directed, and weighted graph. (d) a fully-connected, undirected, and unweighted graph.

Many early studies worked with complex networks modeled using the random network theory (Erdös & Rényi 1960). Random networks assumed that each pair of nodes in the network had a uniformaly random chance of being connected. The result of the model was a homogeneous network with a consistent number of links for each node. More recent studies have deviated heavily from the random network models in replication of structures found on the internet (Albert et al. 1999). This alternative model, "scale-free network", is much less homogeneous. Rather than sharing uniformly similar degrees nodes are often interconnected by a handful of "hub" nodes with a large number of connections (Fig 9 a). These scale-free networks better model situations in which new nodes are added by attaching themselves to existing nodes, so called "preferential attachment". As this happens, nodes with higher connections are more preferentially targeted by the new nodes and these higher degree nodes gain more connections more rapidly (Strogatz 2001).

A final common structure in social networks is a community structure. This structure defined by subsets of nodes with relatively high intra-set connectedness but low inter-set connectedness (Fig 9 b). Depending on what entities are being modeled, communities in a network can represent social groupings by interest or background, citation communities could represent academic fields or niche areas of study, or in a biological model communities could represent metabolic cycles (Girvan & Newman 2002). These community structures can also be found in a hierarchical fashion in which the communities join together into metacommunities which can further join together on a larger network. Each of these network types and characteristics can be heavily influenced by the entities being modeled but can also affect the sorts of analysis which can be done onto the network itself. Because of this, we explored multiple ways of creating the network and examined the sorts of models each process would create.

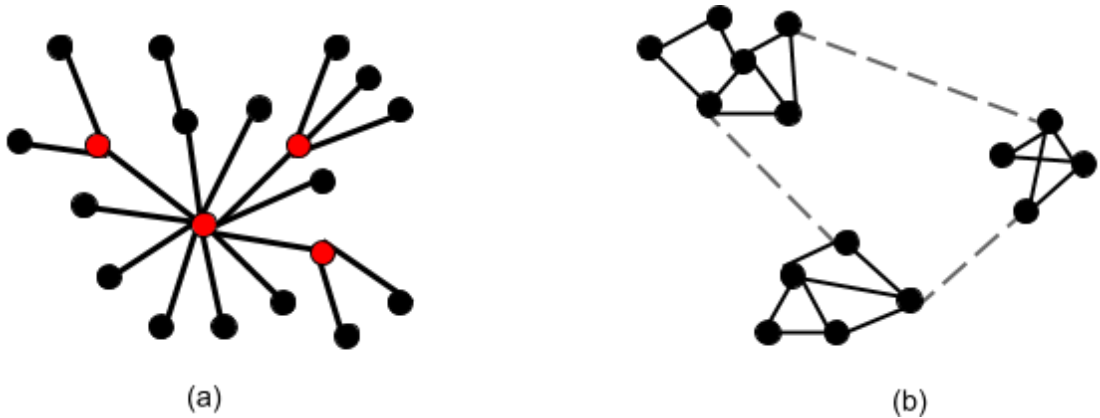

Figure 9. Complex Graph Structures: (a) representation of a scale-free network with hubs in red. (b) representation of three communities in a social network model.

## Our Network Model

Due to the microscopic size of mitochondria and the resolution of our image data, individual mitochondria could not be resolved; only collections of mitochondria could be modelled. Accordingly, we modeled communities of mitochondria as nodes as opposed to each individual mitochondrion.

An early approach to identifying nodes in our data was to find connected components within the structure. Because areas of high flourescence implied a high concentration of mitochondria, a connected component approach would split areas of high concentration of mitochondria along boundaries where the concentration is lower. This process fit a new threshold through Otsu's method to the single cell image, resulting in a binary image, and then applied a connected component labeling algorithm (Wu et al.

2004) to this new binary image. The connected component algorithm labels all pixels which form a contiguous structure in the binary image with one value. In Figure 10 we can see these separate components labeled by color. Each of these separate components would be considered a node in our network. Initially this resulted in a small number of large communities which we would have then considered nodes in our network model (Fig 10, left). However, we wanted more homogeneity between the nodes themselves in order to better model the connection between them. In order to break these larger communities down we applied a stricter threshold in the areas of the larger components in our initial connected component label. This created more boundaries within the larger communities. We then relabeled the broken-up components. This process created more homogenous nodes of similarly sized communities of mitochondria (Fig 10, right). When the threshold removed area from the image, data is lost. This removal of data in node boundaries is the main drawback of the connected component approach. This data removal is exacerbated by the process which further separates the larger components, as cell area is removed by the second thresholding step (Fig 10). Another significant issue in the connected components approach is the difficulty in linking the nodes temporally. Each frame is processed individually when using this method, resulting in a varying number and positioning of nodes over time which are difficult to link to allow for temporal analysis downstream.

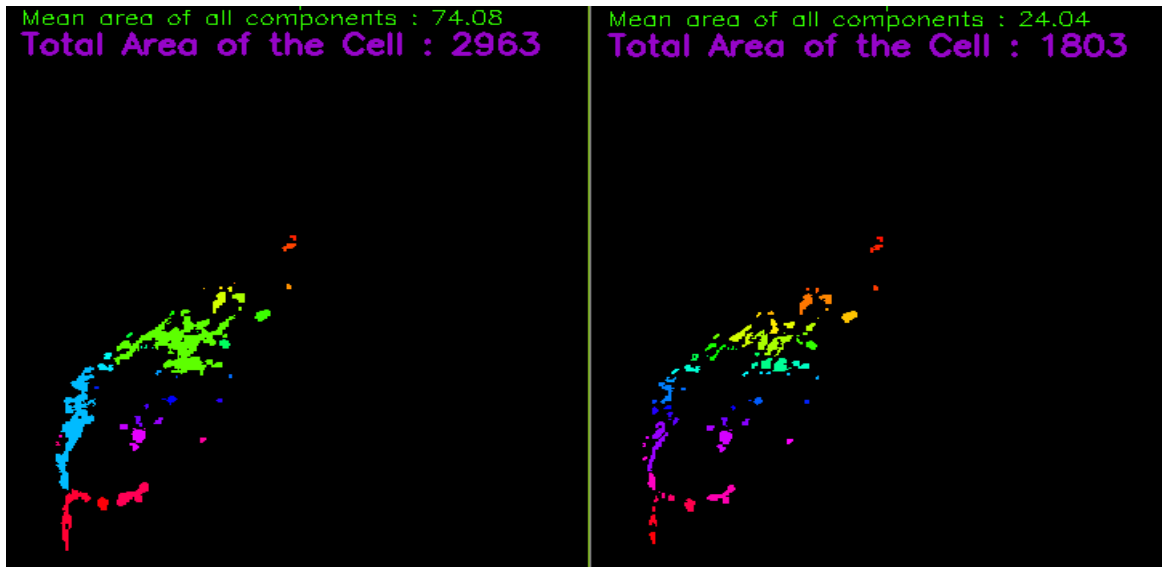

Figure 10. Connected Component Results: (Left): the initial connected components approach, notice large green, blue, and red components. (Right): connected component approach after large components are broken up. Notice how smaller components are unchanging in shape in size while the large components are broken into more similarly sized nodes. Also notice the drastic decrease in area of the cell.

We later moved towards modeling the protein structure by making use of a Gaussian mixture model (GMM). A recent study used a GMM to model the diffuse structure of mitochondria in three-dimensional imaging. They fit a spherical GMM using seeds at each intensity local maxima after smoothing the image with a Gaussian filter. This process localized the mitochondria in relation to the distance to the nucleus and cell surface (Ruan et al. 2018). This confirmed our decision that a GMM would be an ideal way to model the mitochondrial patterns as nodes in a network. Unlike the connected components approach, no areas of the segmented image are lost when using the GMM as all pixels are considered by each Gaussian in the mixture model. Another advantage is that the means and covariances of the model components represented two critical features of a social network: the individual nodes (means), and the nodes' relative relationships to each other (covariances). The GMM could be trained separately for each distinct cell, permitting the model parameters to evolve over the course of the videos to capture the dynamic cell morphologies.

A GMM fits a mixture of Gaussian distributions to a set of data. The model assumes that all of the data points being fit are generated from a mixture of a finite number of Gaussian distributions. For our purposes we are modeling the spatial communities of mitochondria as two dimensional Gaussian distributions, the combined mixture of which gives rise to the diffuse mitochondrial pattern of the cell. The GMM is fit through an expectation-maximization (EM) algorithm. EM is an iterative process which computes a probability for each point belonging to each Gaussian component of the mixture model. The gaussian parameters are then tweaked to maximize the likelihood of the underlying data points given the model. When repeated this process will eventually converge to a local optimum.

---
**Algorithm 2** Fit Gaussian Mixture Model to Video

**Input:** A video $V \in R^{h \times w \times f}$
**Output:** A set of means for each frame $means \in R^{f \times n \times 2}$; A set of covariance matrices for each frame $covars \in R^{f \times n \times 2 \times 2}$

1: **function** GETNODES($V$)
2:    $means, covars \leftarrow \emptyset$    ▷ initialize output with empty set
3:    $MU, CV \leftarrow$ GETSEEDPARAMS($V_0$)    ▷ create seed parameters for video
4:    **for** $i = 0$ to $f$ **do**    ▷ for each frame in video generate nodes
5:       $X \leftarrow$ PREPROCESS($V_i$)    ▷ preprocess the current frame
6:       $MU, CV \leftarrow$ FITGMM($X, MU, CV$)    ▷ fit a GMM to the frame
7:       $means \leftarrow$ APPEND($means, MU$)    ▷ save frame's means for output
8:       $covars \leftarrow$ APPEND($covars, CV$)    ▷ save frame's covars for output
9:    **return** $means, covars$    ▷ return each node for each frame
---

Figure 11. GMM Application Pseudocode

Before fitting the GMM we followed several preprocessing steps to lessen noise in the input data and create a deterministic fitting process. First, we ran a Gaussian smoothing filter to minimize or eliminate artifacts in the video images. With the image smoothed we normalized the grayscale pixel intensities to sum to one, converting the image to a 2D probability distribution. We repeated this process for each frame (Fig 12).

Next, we needed to find seeds for the Gaussian components for each video. Because each component is intended to model a community of mitochondria in the cell, we treat each local maximum in the image as the initial means for our Gaussian components (Fig 12). Our covariances were initialized as 2x2 diagonal matrices containing the variance of the mean's 8 neighbors. These initial parameters seeded the GMM to fit to the first frame. For each subsequent frame, the previous frame's GMM parameters initialized the fitting process (Fig 11). The learned GMM components parameterized the structure of the mitochondrial pattern. The learned components are nodes in the final network. This process allowed for the network structure to be learned purely from the mitochondrial topology (Fig 13).

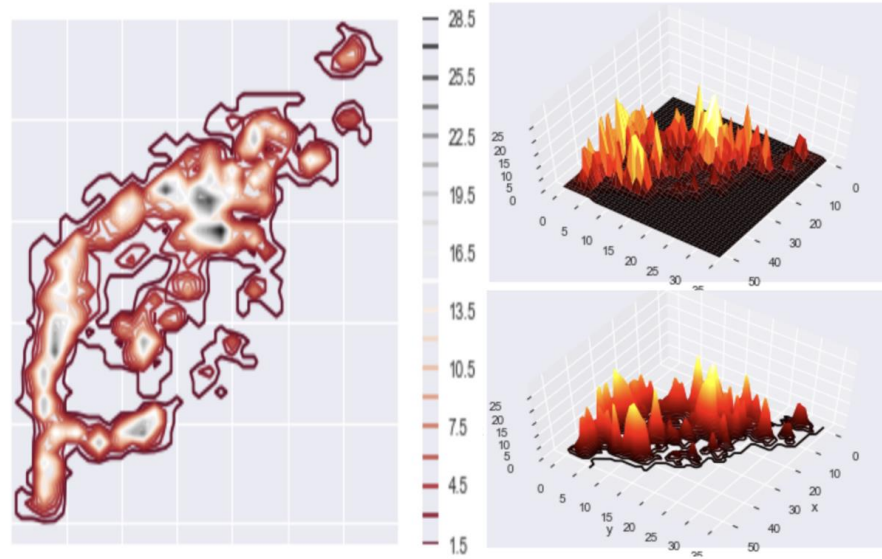

Figure 12. GMM Initialization Contour: (Left) a 2D probability representation of the intensity of a sample cell. (Top Right) the Intensity map of the image in a 3D representation. (Bottom Right) the 3D contour of the same cell

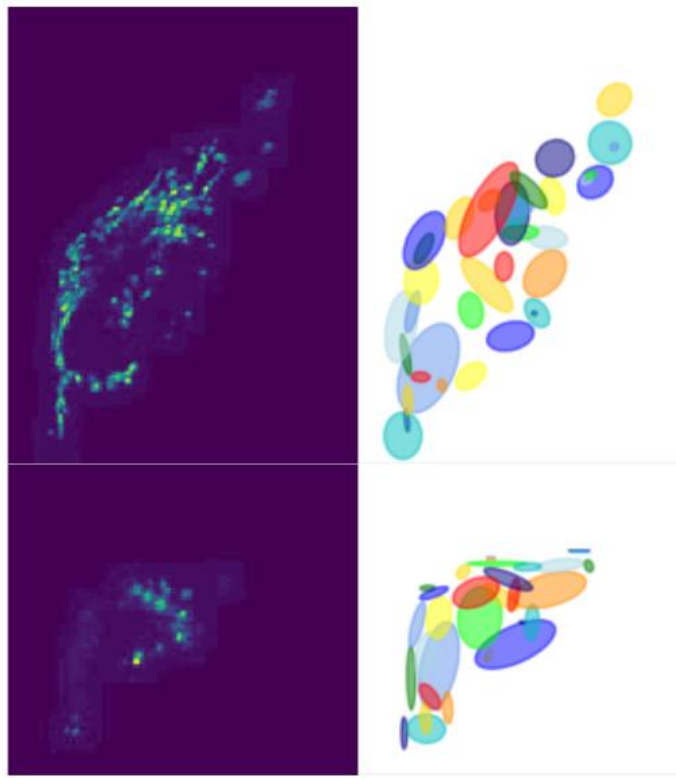

Figure 13. GMM Results: A cell (Left) and the nodes (Right) as generated by a Gaussian mixture model for the first (Top) and last (Bottom) frames of a video showing a cell fragmented by LLO.

To connect the nodes (our GMM components) with weighted edges, we explored multiple approaches that balanced realistically encapsulating the underlying biology (i.e., did not create strong connections between weakly correlated objects) and computational tractability. Initially, we chose a manual distance threshold and used this as the "neighborhood size" for the radial-basis function, a common connection weighting metric that varies smoothly from 0 (not connected) to 1 (fully connected) and is a function of the Euclidean distance between the two nodes, weighted by the neighborhood size. This created a partially connected, undirected graph. However, we wanted to avoid a manually set threshold.

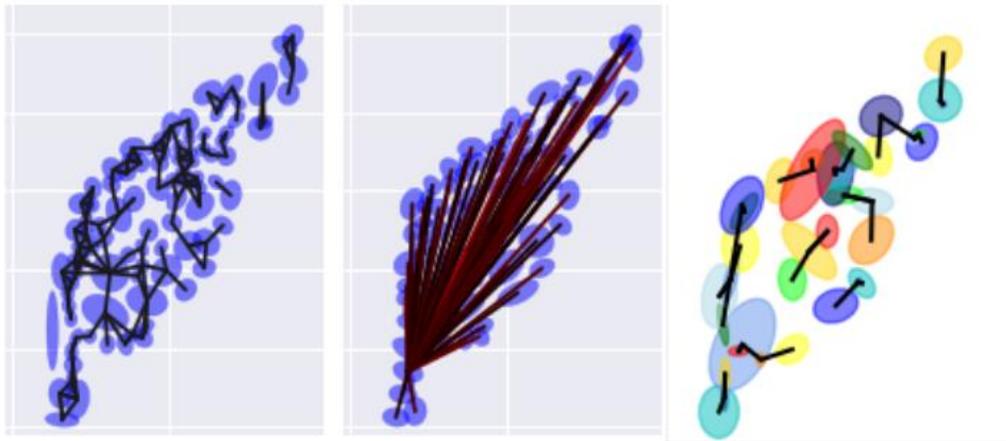

Figure 14. Edge Illustrations: (Left) A partially connected network with binary connections. (Center) A single node's weighted connection within a fully connected graph. (Right) A the strongest connection of each node as determined by our probability-based affinity function

A second attempt to make this process more data-driven was to replace the manually-crafted neighborhood size with the Gaussian covariance in the direction of the node to be connected (Fig 14, mid). In both cases, to avoid fully-connected graphs and induce some sparsity, we set a hard threshold on the maximum distance between nodes to connect (Fig 14, left). However, since these metrics were based in Euclidian distance, they would be dependent on the microscope scale during image capture. They also didn't take the directionality of the components into account when deriving the weights themselves.

To address these shortcomings with determining network connectivity, we computed weights between the Gaussian components using probability. This not only accounted for the anisotropy in the covariance of the Gaussian components, but also captured the asymmetry between components: by decoupling the direct link to Euclidean distance, the connections could instead be weighted by how probable the location of the node under consideration was (Fig 14, right). This resulted in an asymmetric graph matrix, but it more accurately reflected the dynamics of the underlying biology, captured the relationships between nodes in a more intuitive metric, and was entirely data driven with no hand-crafted thresholds.

This newer approach nonetheless only offered a pointwise comparison between the means of each node. This meant that nodes could have a weak connection if the probability of the mean point of one is low in the other, despite the component covariances having an overlapping region in the outlying region along the principle axis. To move away from these pointwise metrics and determine a weighted connection that considered the complete distribution, we used divergence metrics. The first divergence metric we used was Kullback-Leibler (KL) divergence. KL divergence is also known as relative entropy and is used as a measure of similarity between two probability density functions $f(x)$ and $g(x)$.

$$D(f||g) = \int f(x) log \frac{f(x)}{g(x)} dx \qquad (1)$$

The KL divergence is a popular statistical metric due to its properties of self-similarity, self-identification, and positivity. As two distribution become more identical, their KL divergence nears zero. However, *D(f // g) = 0* only if *f = g*. We used the KL divergence as the weight of the connection between two nodes, creating a directed network as KL divergence is asymmetric. Biologically speaking, there is little evidence to prefer a directed graph structure over an undirected one. Yet, with a cell's general Brownian behavior, the undirected structure seems more analogous and flexible. In order to create an undirected network, we used a related divergence method, Jensen-Shannon (JS) divergence. JS divergence is a metric based on KL divergence, with two notable differences. Firstly, JS divergence is always finite, which allows for more well-behaved networks for further analysis. More importantly, JS divergence is a symmetric metric (2) which, when used as weights in our social network, gives us the undirected network structure which we desire to model the mitochondrial proteins.

$$JSD(f||g) = \frac{1}{2}(D(f||M) + D(g||M)) \; where \; M = \frac{1}{2}(f+g) \qquad (2)$$

With this edge creation method in place we were finally able to generate network structures for mitochondrial proteins resulting in a fully connected, undirected, weighted graph with structure derived from the spatial distribution of mitochondria. Computing this model for each frame allows the model to evolving over time to follow the dynamic pattern. This process served as a basis by which the networks could be created, with some elements of the graph structure becoming more cemented as we analyzed the resulting networks.

CHAPTER 4

MODEL ANALYSIS

As we built our early networks using the described GMM method with probability-based connection weights for each cell under varying conditions (control/ wildtype and LLO), we observed qualitative systemic differences in the learned model parameters that separated these conditions. Interestingly, as the mitochondria fragmented (i.e., LLO), the GMM components became more strongly connected, not less (Fig 15). We attributed this to a misinformed intuition: as the mitochondria fragmented and the underlying probability density function became more uniform, the GMM components likewise became more uniform, resulting in a more uniformly connected network. The overall number of connections also increased in thresholded networks, as the cells tended to collapse at the same time as mitochondrial fragmentation, resulting in the same number of GMM components spatially co-locating in a much smaller space, effectively "forcing" connections by virtue of proximity. By comparison, the control cell showed much less variation in the distribution of network connectivity and edge weights over time; this reflects a relatively stable social network, unperturbed by external stimuli.

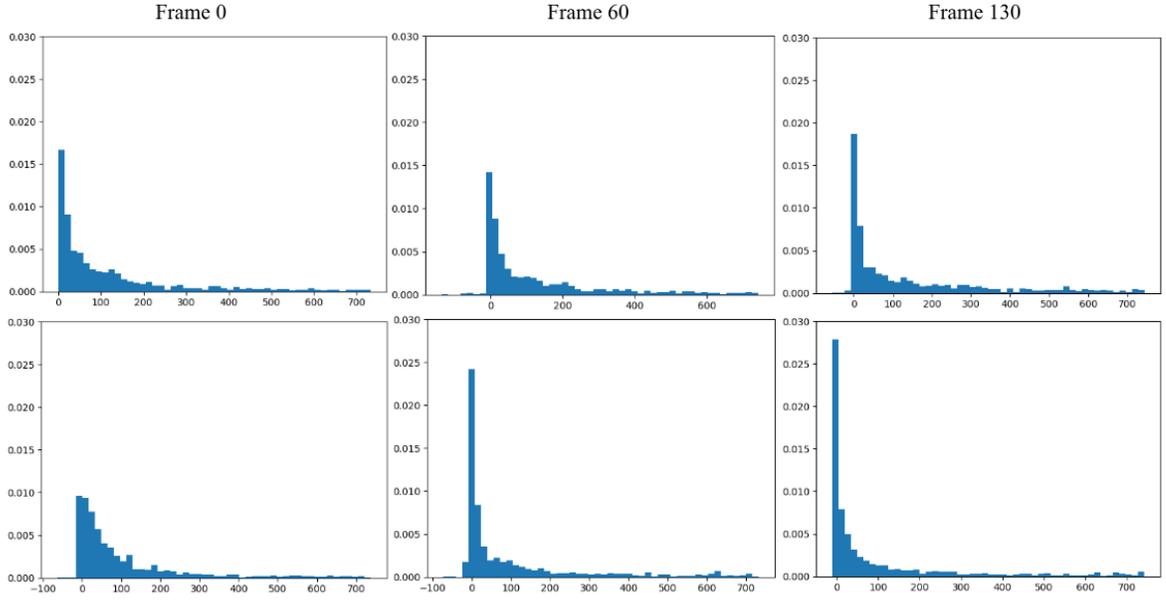

Figure 15. Affinity Distribution: A series of distribution plots of the negative log of values found in six affinity tables developed using the model learned at an early, middle, and late video frame. (Top) The tables generated from a control cell which show little variation in distribution. (Bottom) The tables generated from the LLO cell which shows a surprising increase in connectivity over time as the cell fragments as the negative log approaches 0 when the probability nears 1.

The next step was to develop a temporal model of the GMM component evolution in terms of the social network topology. We created a series of graph Laplacians, one for each frame, and observed how the Laplacians changed through spectral decomposition. By virtue of the Jensen-Shannon divergence, the Laplacians had to be modified. In order to correct for this, we applied a heat kernel (3) to transform this JS matrix into a similarity matrix so that the network would be suited for spectral decomposition.

$$S_i = e^{\frac{-\beta \times A_i}{\sigma}} \text{ where } \beta \text{ is free and } \sigma \text{ is the standard deviation of } A \quad (3)$$

Once each network's JS matrix was converted to similarities using the heat kernel, the graph Laplacians for each frame could be found. We then applied an eigensolver to decompose each Laplacian into eigenvectors and eigenvalues.

This spectral decomposition process allowed us to view the spectrum of each frame's Laplacian. We then plot the ten principal eigenvalues (ten with the highest value) in the spectrum together over time. This allowed us to observe qualitative differences in the temporal behavior of cells in each of the varied conditions (control/wild type, LLO, mdivi).

The wild type cell, whose morphology offers little variance over time, exhibited stability in its spectrum (Fig 16 center). This is exhibited both in the long term as values neither increase nor decrease greatly over time, as well as in the short term with minor fluctuations in most of the eigenvalues throughout the timeframe of the video. This is visible in the y-scale difference in figure 16 which only increases by a half unit each tick while the others are twice as large. The control cell's spectrum also suggested a more well-behaved system due to the lower value of the components. The most principal component has an eigenvalue ranging from 3.0 to 3.5 while the other cells have principal components >10.

The Mdivi cell, unlike the wild type cell, showed instability over time. There was a relatively large decrease in all eigenvalues from frame 0 to frame 120 (Figure 16, right) which illustrated a significant change in morphology. Much of this decrease occurs in the first 40 to 50 frames. However, even after the eigenvalues level off around frame 50 there is still significant short-term variance (Figure 16, right). This short-term variance suggested a relatively uniform probability distribution, with which the GMM struggles to find a consistent mixture between consecutive frames. This uniform probability distribution describes the hyperfused morphology induced by mdivi.

The LLO cell's spectral plot was the most unique. The fragmentation period was clearly visible in frames 40-60 (Fig 16, left). This period was illustrated by a significant and violent shift in eigenvalues as the model reshapes itself to fit to the fragmented morphology. The most principal components had consistent values after the fragmentation. However, the less principal eigenvalues showed significant short-term fluctuation like that of the mdivi. While the result is similar, the underlying cause is quite different. The hyper-fused cell's spectral fluctuations were the result of a uniform probability distribution, while the fragmented cell's distribution is far from uniform. The cause for fluctuation in the less principal components was the drastic decrease spatial area of mitochondrial protein after fragmentation (Figure 13). After fragmentation there were fewer communities than nodes and nodes were forced to fit to background noise. The GMM struggled to create consistent components for these regions resulting in short-term fluctuation of the eigenvalues. However, since a handful of nodes were able to fit to the remaining communities our most principal eigenvalues remained relatively stable after fluctuation.

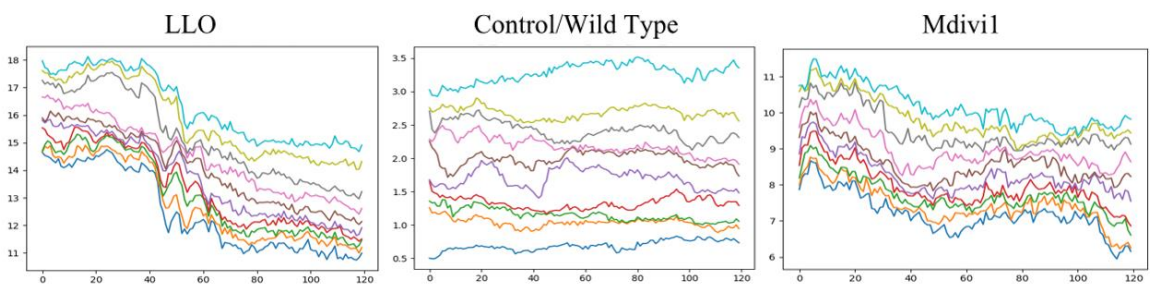

Figure 16. Temporal Eigenvalue Analysis: The eigenvalues of the first 120 Laplacians for an LLO, Control, and Mdivi video respectively.

CHAPTER 5

NEXT STEPS AND CONCLUSION

Model Limitations and Next Steps

The next steps in the analysis are to observe the change in the series of Laplacians derived from each video as a function of Laplacian gradients. This would highlight specific portions of the social networks that covary over space and time; in other words, it would provide insight into the coordinated fragmentation or hyper-fusion of the mitochondria in response to the provided stimulus. These features could then be incorporated into a broader supervised learning pipeline to distinguish patterns and discern the effects of an unknown stimulant (e.g., drug discovery), or an unsupervised learning pipeline to identify all observed mitochondrial phenotypes.

Additional methods of analyzing the graph structure of the social network would help determine specific phenotypic changes induced by certain stimuli. Classic graph metrics such as connectivity, cliques, and eigenvector centrality would help to precisely measure the global effects of certain stimuli on the mitochondria. Other algorithms, such as PageRank for global network analysis from local phenomena would provide intuition into the local changes in mitochondrial phenotype responsible for inducing the global structure. These metrics would be invaluable for characterizing certain specific cell-wide or even organism-wide conditions.

Further improvements can also be made to the network generation pipeline itself. We began working to incorporate a single uniform component into the overall GMM to provide a robust method of accounting for background noise in the form of a learned, data-driven threshold.

Another shortcoming of the current GMM approach is a "node collapse" in some LLO samples. When the protein pattern fragments to a certain extent, some nodes will collapse because of a lack of probability mass in the area near the node. In our analysis, we removed collapsed nodes from the video, causing a loss of data but preventing these collapses from corrupting the remaining data. While these nodes were mostly small and accounted for small amounts of the overall protein structure, any data loss should be avoided in the modeling process if possible. An modification to the pipeline which could potentially both fix this issue and improve the model's capability to mimic the protein mass is a flexible number of components. A large amount of information in temporal social network analysis comes from the addition and removal of nodes as entities appear and leave the community. Currently our model does not allow for the addition of nodes to represent fission of mitochondrial communities or the removal of nodes to represent fusion. By adding this, a fragmentation period in LLO cells could be much more easily quantified by a change in the number of components as the communities dissipate.

Another shortcoming of our pipeline was the manual initialization. With deep learning models at the cutting edge of semantic segmentation in biomedical images, the masks created in this study for the thousands of frames microscopy images could serve as training data for fully autonomous segmentation of mitochondrial protein structures. This would allow this pipeline to be more quickly and flexibly applied to larger sets of data while also allowing an expansion in the area of deep models for autonomous semantic segmentation.

## Conclusion

In this paper, we have presented a proof-of-concept for parameterizing and modeling spatiotemporal changes in diffuse subcellular protein patterns using GMMs. Our pipeline is intended to be flexible for future use in modeling other similarly diffuse subcellular structures such as actin tubules, or for the purposes of other network-based analytics which necessitate alternate network structures (directed/undirected, dense/sparse, etc.). We have presented how the learned parameters of the GMM can be updated to account for changing biological phenotypes, and how these parameters can then be used to induce a social network of interacting nodes. Finally, we show how the properties of the social network can be interpreted to provide biological insights, particularly how the underlying system may be responding to some kind of stimulus. This pipeline has potential implications in fundamental biology and translational biomedicine; we aim to complete our analysis package and release it as open source for the research community to use in the near future.